\documentstyle[times,pramana,epsf,floats]{ias}
\newcommand{\s}{\rm}

\newcommand{\ra}{\rightarrow}

\newcommand{\be}{\begin{equation}}
\newcommand{\ee}{\end{equation}}

\newcommand{\bea}{\begin{eqnarray}}
\newcommand{\eea}{\end{eqnarray}}
\newcommand{\bef}{\begin{figure}}
\newcommand{\eef}{\end{figure}}

\newcommand{\lgl}{\langle}
\newcommand{\rgl}{\rangle}

\begin{document}

\mark{{Photons from Quark Gluon Plasma and Hot Hadronic Matter}{Jan-e Alam}}

\title{Photons from Quark Gluon Plasma and Hot Hadronic Matter}

\author{Jan-e Alam}

\address{Variable Energy Cyclotron Centre, 1/AF Bidhan Nagar Kolkata 70064,
India 
}
\keywords{Photons, QGP, Phase Transition, Spectral Shift, Hydrodynamics}
\pacs{2.0}

\abstract{The productions of real photons 
from quark gluon plasma and hot hadronic matter
formed after the nucleus - nucleus collisions at
ultra-relativistic energies are discussed.
The effects of the spectral shift of the hadrons at
finite temperature on the production of photons
are investigated. On the basis of the present
analysis it is shown that the photon spectra measured
by WA98 collaboration in Pb + Pb collisions at CERN
SPS energies can be explained by both QGP as well as hadronic
initial states if the spectral shift of hadrons at
finite temperature is taken into account. 
Several other works on the analysis of WA98 photon data
have also been briefly discussed. 
}

\maketitle
\section{Introduction}
Calculations based on the 
QCD renormalization group  
predicts that strongly interacting systems at very high 
density and/or temperature
are composed of weakly interacting quarks and gluons
~\cite{collins} due to 
asymptotic freedom and the Debye screening of colour charge. 
On the other hand at low temperature and
density the quarks and gluons are confined within the hadrons.
Therefore, a phase transition is  expected to take place
at an intermediate value of temperature and/or density.
In fact such a transition, from  
hadronic matter to Quark Gluon Plasma (QGP) is actually 
observed in lattice QCD
numerical simulations \cite{latt} at high temperature. 
One expects that ultra-relativistic heavy ion 
collisions (URHIC) at CERN/SPS, BNL/RHIC and CERN/LHC
might create conditions conducive 
for the formation and study 
of QGP ~\cite{qm01}. 
Among various signatures of QGP, photons and
dileptons are known to be advantageous as these signals 
probe the entire
volume  of  the  plasma,  with little interaction and thus, are
better markers of the space-time history of the evolving matter
~\cite{emprobe}.

The  aim of the present work is 
to contrast the real photon emission rate 
from the following two nuclear collision scenarios:
\begin{center}
{\bf (1) A\,+A\,$\ra$QGP$\ra$Mixed Phase$\ra$Hadronic Phase\\
or\\
(2) A\,+\,A\,$\ra$Hadronic Phase},
\end{center}
by taking into account the finite temperature effects on the hadronic 
masses and decay widths. The main sources of photons from URHIC 
are: (a) the decay of mesons (mainly $\pi^0$ and $\eta$), 
(b) thermal source, either QGP and/or the hadronic reactions and hadronic 
decays in the thermal medium depending on which of the scenarios (1) or (2) 
realized after the collisions and 
(c) hard scattering of the partons embedded in the nucleons of
the colliding nuclei in the very early stage of the collisions.
However, the transverse momentum ($p_T$) spectra of 
single photons presented by WA98 collaboration~\cite{wa98} 
does not contain the contributions 
from the decays of mesons.
Therefore, in the present article we will consider photons
from (b) and (c) only.
In the scenario (1) the pre-requisite for the QGP diagnostics (by measuring
photon spectra) is to estimate the photon yield from hadronic sources 
and hard collisions of partons in the initial stage.   

In the next section we discuss  
the hard photon production from nucleus-nucleus 
collisions at SPS energies. Thermal photon emission
from the QGP and hadronic matter are discussed in section 3.
Hadronic properties at non-zero temperature is
presented in section 4. 
Section 5 is devoted to discuss
the space time evolution and in section 6 we present
our results and discussions.

\section{Hard QCD Photons}

The hard QCD photons are estimated using perturbative QCD as
\be
E\frac{dN}{d^3p}=T_{AA}(b=0)\,\,E\frac{d\sigma_{pp}}{d^3p}
\ee
where $T_{AA}(b)$ is the nuclear thickness at impact parameter $b$. 
Its value at $b\sim 3.2$ fm, corresponding to 
the most central event of WA98 experiment is $\sim\,220$/fm$^2$. 
$\sigma_{pp}$ includes the $pp$ cross-section for
Compton and annihilation processes among the partons. At SPS energies one
should include the effects of intrinsic $k_T$ distribution
of partons~\cite{ownes} (due to finite size of the
nucleons). 
This leads to substantial enhancement 
in the photon spectra~\cite{cywong}. 
In practice such an effect is implemented by multiplying
each of the parton distribution functions appearing in the right hand side 
of the above equation by a Gaussian function of the type $f(k_{T})
=\exp[-k_T^2/\lgl k_T^2\rgl]/\pi\lgl k_T^2\rgl$ and integrating over
$d^2k_T$.
We use CTEQ5M partons~\cite{cteq} 
and $\lgl k_T^2\rgl=0.9$ GeV$^2$ for
evaluating the hard QCD photons. 
The energy at the centre of mass of Pb + Pb collisions at SPS
is 17.3 GeV. Experimental data on hard
photons does not exist at this energy. 
Therefore, the ``data'' at $\sqrt{s}=17.3$
GeV is obtained from the data at $\sqrt{s}=19.4$ GeV of the E704
collaboration ~\cite{e704}
by using the scaling relation: 
 $Ed\sigma/d^3p_\gamma\mid_{h_a+h_b\,\ra\,X+\gamma}=f(x_T=2p_T/\sqrt{s})/s^2$,
for the hadronic process, $h_a+h_b\ra X+\gamma$~\cite{tf}.
However, such a scaling may be spoiled in perturbative QCD due to
the momentum dependence of the 
strong coupling, $\alpha_s$ and from the scaling violation
of structure functions, resulting  
in faster decrease of the cross section than
$1/s^2$. Therefore, the data at $\sqrt{s}=17.3$ GeV
obtained by using the above scaling gives a conservative
estimate of the prompt photon contributions.  
The photons from hard QCD processes
have been used to reproduce the scaled p-p data of E704 collaboration. 
The higher order effects has been
taken into account through a K-factor $\sim 2$. 
Now the question is, can we say that an enhanced production in 
A-A collisions compared to p-p will presumably mark the presence of a 
thermal source? Not necessarily, because in nucleus-nucleus
collisions there may be enhancement in the high $p_T$
part of the photon spectra due to various 
nuclear effects, e.g., Cronin effects.
However, the good news is that even if we take into account
the transverse momentum broadening  due to the finite size
of the nucleon as well as due to the Cronin effects 
then we find that the theoretical yield is less than the experimental 
value (WA98 data) for $1.5<p_T$(GeV)$<2.5$~\cite{dumitru,proy}. 
Clearly indicating the presence of thermal source.  
What is the nature of this thermal source
which can reproduce the WA98 data?  We 
will discuss this issue in the next section.

\section{Thermal photons from QGP and hadronic matter}
Since quarks are electrically charged their interactions
at the thermal bath will produced photons we are looking for.
The quark - anti-quark annihilation and QCD compton
are the dominant processes for the production of photon
from a thermalized system of quarks and gluons~\cite{kapusta}.
However, it has been
shown~\cite{aurenche} (see also~\cite{st,dd}) 
that the two-loop contribution leading to
bremsstrahlung and $q\bar q$ annihilation with scattering is of
the same order as the lowest order processes. The total rate of emission
(upto two loops) per unit four-volume at temperature $T$ is given by
\bea
E\frac{dR}{d^3p}&=&\frac{5}{9}\,\frac{\alpha\alpha_s}{2\pi^2}\,\,
\exp(-E/T)
\,\left[\ln\left(\frac{2.912\,E}{g^2\,T}\right)\right.
\nonumber\\
&&\left.+4\frac{(J_T-J_L)}{\pi^3}
\{\ln2+\frac{E}{3T}\}\right]
\label{edr}
\eea
where $J_T\simeq4.45$ and $J_L\simeq-4.26$. 
The QCD coupling, `$g$' is given by,
\be
g^2/4\pi\equiv\alpha_s=\frac{6\pi}{(33-2n_f)\ln(\kappa\,T/T_c)}
\label{strong}
\ee
where $n_f$ is the number of quark flavours and $\kappa=8$~\cite{fk}.
The photon emission rate in Eq.~(\ref{edr}) is evaluated  in the
Hard Thermal Loop (HTL) approximation. However, the HTL approximation
is not valid at SPS energies  where an initial temperature about few 
hundred MeV
may be realized. Because the HTL resummation is valid for $g<<1$ whereas the 
value of $g$ obtained from Eq.~(\ref{strong}) is $\sim 2$ at $T\sim 200$ MeV.
At present it is unclear whether the rate in Eq.~(\ref{edr}) is valid for
such a large value of $g$ or not. 
It would be rather difficult to make any firm conclusion 
from the results where Eq.~(\ref{edr}) is used at SPS energies.
Keeping this reservation in mind 
we will use Eq.~(\ref{edr}) to evaluate the photon yield from QGP. 
However, we will see below that
within the present framework the space time integrated photon yield
from quark matter is less than that from hadronic
matter due to the smaller life time of the QGP phase as a
result of a moderate value of the initial temperature considered here. 
Therefore,
the total thermal photon yield remains largely unaffected 
even in a scenario where QGP is formed in the initial state. 

The photon yield from the hadronic matter (HM)
(see first of~\cite{kapusta}), 
is evaluated from the reactions,
$\pi\,\rho\,\ra\, \pi\,\gamma$, 
$\pi\,\pi\,\ra\, \rho\,\gamma$, $\pi\,\pi\,\ra\, \eta\,\gamma$, 
$\pi\,\eta\,\ra\, \pi\,\gamma$ and the decays $\rho\,\ra\,\pi\,\pi\,\gamma$
and $\omega\,\ra\,\pi\,\gamma$.
We refer to Refs.~\cite{npas} for the invariant amplitudes of these
processes.
Photon production due to the process $\pi\,\rho\,\ra\,a_1\,\ra\,\pi\,\gamma$
is also considered here. 

\section{Hadronic properties at $T\,\neq\,0$}

It has been emphasized that the hadronic properties will be 
modified due to its interactions with the particles in the
thermal bath. As a consequence of the change in the 
properties of hadrons the static photon emission rates as well as 
the Equation of State (EOS) of the evolving matter will also change 
in a non-trivial way ~\cite{annals}. Broadly two kinds of medium 
modifications of hadrons are expected: (i) shift in the pole of the 
spectral function without any broadening  and 
(ii) broadening of the spectral function but pole does not shift.
In Ref.~\cite{annals} the effects of spectral changes of hadrons
on the electromagnetic probes is studied in detail.
It was observed that the gauged linear and non-linear  
sigma models and the model with hidden local symmetry do
 not show any appreciable effect on photon emissions.
 In the Walecka model, the universal scaling
 hypothesis for the hadronic  masses (except pseudo-scalar) 
has been seen to enhance photon emission.

Both (i) and (ii) can reproduce the enhancement of 
the dilepton yield in the
low invariant mass region, however, 
the photon yield is largely unaffected
by (ii) since the spectral function is smeared out.
Therefore, a simultaneous measurements of single photon and lepton
pairs is important to shed light on the in-medium effects of hadrons.
According to the universal scaling scenario~\cite{brpr} the in-medium 
quantities (denoted by $*$) at finite $T$ is parametrized as
\be
{m_{V}^* \over m_{V}}  = 
{f_{V}^* \over f_{V}} = 
{\omega_{0}^* \over \omega_{0}}  =
 \left( 1 - {T^2 \over T_c^2} \right) ^{\lambda},
\label{anst}
\ee
where $V$ stands for vector mesons, $f_V$ is the coupling between the vector
meson field and the electromagnetic current and $\omega_0$ is the continuum threshold. Mass of the nucleon varies with temperature
as in Eq.~(\ref{anst}). 
It is to be noted that there is no definite reason to believe that all the in-medium dynamical quantities are dictated by a single exponent $\lambda$.
This is the simplest possible ansatz. 
The effective mass of $a_1$
is estimated by the Weinberg's sum rules~\cite{weinberg}.
 
In the QHD model~\cite{vol16} the effective masses of 
nucleon, $\rho$ and $\omega$ mesons
can be parametrized in the following forms:
\be
M_N^\ast=M_N\left[1-0.0264\left(\frac{T ({\s GeV})}{0.16}\right)^{8.94}\right].
\label{qhdn}
\ee
\bea
m_\rho^\ast&=&m_\rho\left[1-
0.127\left(\frac{T({\s GeV})}{0.16}\right)^{5.24}\right]\nonumber\\
m_\omega^\ast&=&m_\omega\left[1-
0.0438\left(\frac{T({\s GeV})}{0.16}\right)^{7.09}\right].
\label{qhdv}
\eea

In (ii) the change of the width of the $\rho$ meson with temperature is parametrized as,
\be
\Gamma{_\rho}^*=\Gamma_\rho/(1-T^2/T_c^2)
\label{ewidth}
\ee
and the mass remains constant to its vacuum value. 
The results for scenarios (i) and (ii) will be compared with a scenario 
(iii) where both the masses and widths of the hadrons remain fixed to
their vacuum values. 
\begin{figure}[htbp]
\epsfxsize=8cm
\centerline{\epsfbox{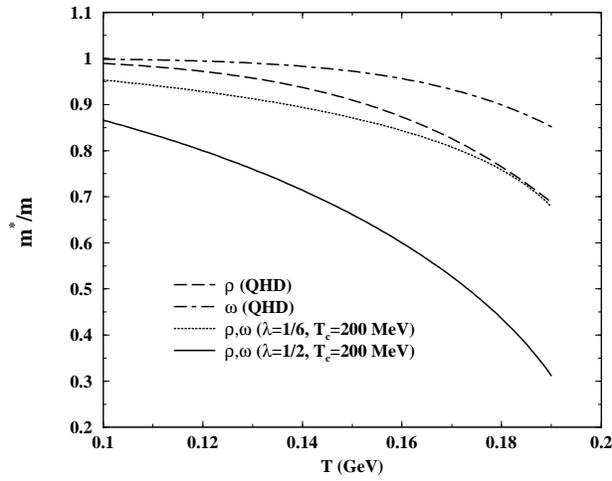}}
\caption{Variation of vector meson masses as a function
of temperature in QHD model and universal scaling scenario.
}
\label{figmass}
\end{figure}


\begin{figure}[htbp]
\epsfxsize=8cm
\centerline{\epsfbox{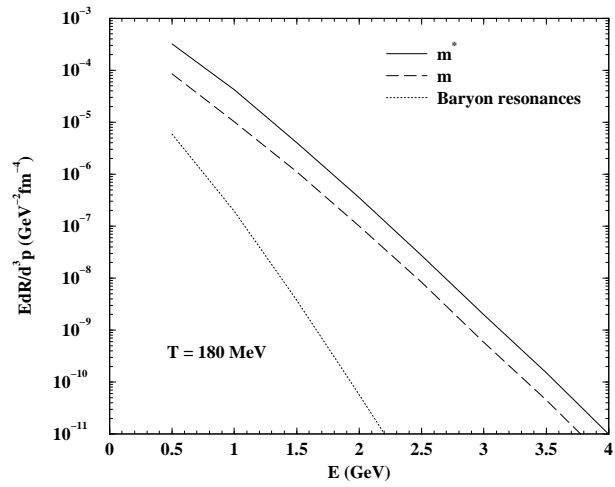}}
\caption{Photon spectra at $T=180$ MeV. Solid (dashed) line indicates 
result when hadronic masses vary according to Eq.~(\protect\ref{anst}) (fixed
at vacuum values). Dotted line shows photon spectra from the decays
of baryonic resonances.
}
\label{figbaryon}
\end{figure}

\section{Space Time Evolution}
It is assumed here that the produced matter reaches a state of thermodynamic
equilibrium after a proper time $\sim$ 1 fm/c~\cite{bj}. In case of a
deconfined matter is produced, it evolves in space and time till
freeze-out undergoing a phase transition to hadronic matter in
the process. The (3+1) dimensional hydrodynamic equations 
have been solved numerically by the relativistic version of the
flux corrected transport algorithm~\cite{hvg}, assuming boost
invariance in the longitudinal direction and cylindrical
symmetry in the transverse plane. 
The initial temperature $T_i$
can be related to the multiplicity of the event, $dN/dy$ by virtue of the
isentropic expansion as~\cite{hwa},
\be
\frac{dN}{dy}=\frac{45\zeta(3)}{2\pi^4}\pi\,R_A^2 4a_k\,T_i^3\tau_i
\label{dnpidy}
\ee
where $R_A$ is the initial radius
of the system, $\tau_i$ is the initial thermalization time and 
$a_k=({\pi^2}/{90})\,g_k$; $g_k$ being the effective degeneracy 
for the phase $k$ (QGP or hadronic matter). 
The bag model EOS is used for the QGP phase. 
$g_H(T)$, the statistical degeneracy of the hadronic phase,
composed of $\pi$, $\rho$, $\omega$, $\eta$, $a_1$ and
nucleons is a temperature dependent quantity in this case 
and plays a crucial  role in the EOS~\cite{annals}. 
As a consequence the square of sound velocity, 
$c_s^{-2}\,=[(T/g_H)(dg_H/dT)+3] < 1/3$, for the hadronic phase,
indicating non-vanishing interactions among the constituents
(see also~\cite{asakawa}).
The hydrodynamic equations have been solved with
initial energy density, $\epsilon(\tau_i,r)$~\cite{hvg}, obtained from $T_i$ 
through the EOS. We use the following relation
for the initial velocity profile which has been 
successfully used to study transverse momentum spectra of hadrons
~\cite{pbm,uh}, 
\be
v_r=v_0\left(\frac{r}{R_A}\right)^\delta
\label{velprofile} 
\ee
Here we took $\delta=1$ 
and the sensitivity of the results on $v_0$ will be 
shown. It is observed that the results 
do not change substantially with reasonable 
variation of the parameter $\delta$ for a given value of $v_0$.  

The space-time integration can also be performed by taking the temperature
profile from the transport model
as~\cite{rapp,li},
\be
T(\tau)=(T_i -T_\infty)e^{-\tau/\tau_0} + T_\infty
\label{trans}
\ee
where $\tau_0=8$ fm/c and $T_\infty=120$ MeV.  
\begin{figure}[htbp]
\epsfxsize=8cm
\centerline{\epsfbox{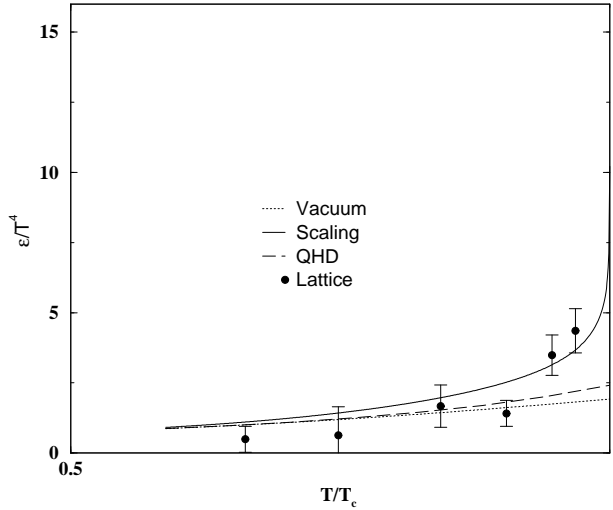}}
\caption{The energy density $\epsilon$ in
the unit of $T^4$ for the equation of state used
in the present work is plotted
as function of temperature ($T$) in the unit 
of the critical temperature, $T_c$. The 
filled circle denotes the lattice results~\protect\cite{latt}.
}
\label{figlat}
\end{figure}


\section{Results and Discussions}
In Fig.~(\ref{figmass}) the effective masses of vector mesons is plotted
as a function of temperature for universal scaling and
QHD model calculations. In QHD the $\rho$ and $\omega$
masses show different behaviour due to their
different coupling strengths with the nucleons in
the thermal bath. 
In Fig.~(\ref{figbaryon}), 
we show the photon emission rate from hadronic matter 
at a temperature, $T=180$ MeV.
The solid and dashed lines correspond to the case 
(i) and (iii) respectively.
The case with large collisional broadening shows no deviation from (iii). 
The increased photon yield at large energy ($E$) is caused by
the enhancement in the Boltzmann factor due to the reduction in meson
(particularly $\rho$) masses. The thermal photon yield with 
hadronic mass variation due to Walecka model or Brown-Rho
scaling~\cite{brpr} ($\lambda=1/6$ in Eq.~(\ref{anst}))
will lie between the solid and dashed curves in Fig.~(\ref{figbaryon}).
It is clearly observed that the contributions from the
decays of baryonic resonances ($N(1520),\,N(1535),\,
N(1440),\,\Delta(1232),\,$ and $\Delta(1620)$) are small
(dotted line). The values of the decay widths, $R\,\ra\,N\,\gamma$,
where $R$ and $N$ denote the baryonic resonances and the nucleon
respectively, are taken from the particle data book. 

We need the equation of state and
the initial condition to solve the hydrodynamic equations.
The effects of the temperature dependent hadronic masses have been
taken into account in the EOS through the effective 
statistical degeneracy~\cite{annals}. In fig.~\ref{figlat}
the temperature dependence obtained for different hadronic 
interactions is compared with the 
lattice QCD calculations~\cite{latt}.
The universal scaling scenario seems to 
reproduce the lattice data reasonably well.

For central collisions of Pb nuclei at 158 AGeV at 
the CERN-SPS, we assume that QGP is produced at $\tau_i$=1 
fm/c which expands 
and undergoes a first order phase
transition to hadronic matter at $T_c$=160 MeV. 
Taking $dN/dy$=700 and
$g_k=g_{QGP}$=37 for a two-flavour QGP, the initial temperature $T_i$
comes out as 196 MeV. In a
first order phase transition one has a mixed phase of coexisting QGP
and hadronic matter which persists till the phase transition is over.
Thereafter the hadronic matter expands, cools and freezes out at a
temperature, $T_f$ and radial velocity,
$v_r^f$.  The sum total of the photon yields from
the QGP phase, the mixed phase and the hadronic phase constitutes the
thermal yield~\cite{japrc}. 
The values of ($T_f, v_r^f$) should in principle be obtained 
from the analysis of hadronic spectra. 
In the present work the value of $T_f$ is taken as 120 MeV 
which reproduces the hadron spectra~\cite{bkp} of NA49 
collaboration~\cite{na49}. The sensitivity of the results on the
value of $v_0$ will be demonstrated below.  

\begin{figure}[htbp]
\epsfxsize=8cm
\centerline{\epsfbox{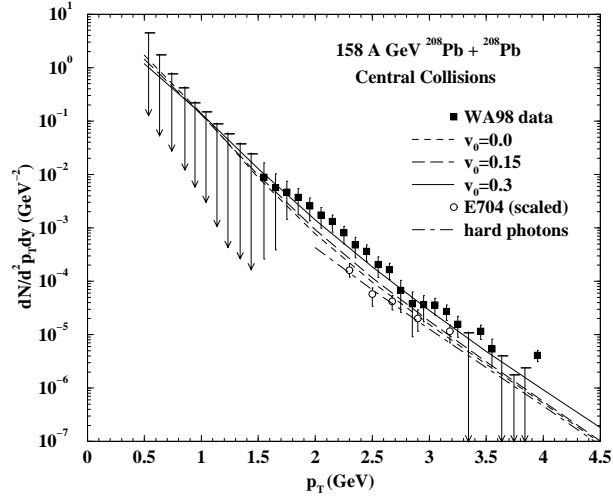}}
\caption{
Total photon yield in Pb + Pb collisions
at 158 A GeV at CERN-SPS. The theoretical
calculations contain hard QCD and thermal 
photons. 
The system is formed in the QGP phase with initial temperature 
$T_i=196$ MeV.  
}
\label{figqgp}
\end{figure}
In case of QGP formation
the {\em thermal photons} contain contributions from 
quark matter (QM $\equiv$ QGP + QGP part of mixed phase) and
hadronic matter (HM $\equiv$ hadronic part of mixed phase +
hadronic phase).  In Fig.~(\ref{figqgp}),
results for the total photon emission is shown for three different values
of the initial transverse velocity with medium effects as in case (i).
All the three curves represent the sum of the thermal and
the prompt photon contribution which includes possible finite $k_T$
effects of the parton distributions. The later, shown separately by
the dot-dashed line also explains the scaled $pp$ data from E704 
experiment~\cite{e704}. 
We observe that the photon spectra for the initial velocity
profile given by Eq.~(\ref{velprofile}) with  $v_0=0.3$ explains
the WA98 data reasonably well. 
It is found that a substantial fraction of the photons come from mixed and 
hadronic phases (hadronic masses
vary with temperature according to Eq.~(\ref{anst})
with $\lambda=1/2$) . The contributions from the QGP phase
is small because of the small life time of the 
QGP ($\sim 1$ fm/c). Therefore, the results shown in this figure
is largely independent on the uncertainties (mentioned above) 
involved in the photon emission rates from QGP. 
    
\begin{figure}[htbp]
\epsfxsize=8cm
\centerline{\epsfbox{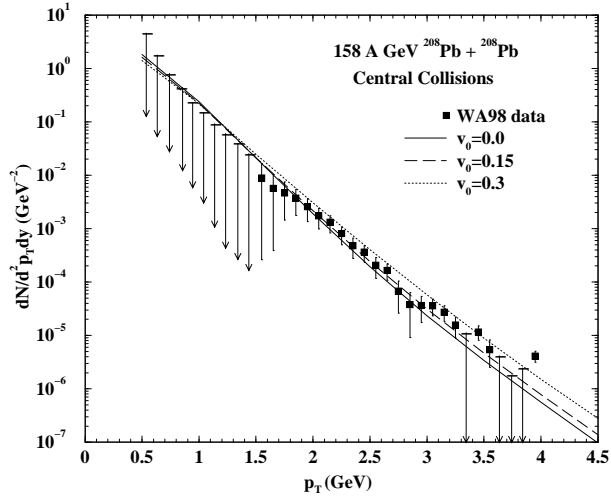}}
\caption{Same as Fig.~(\protect{\ref{figqgp}}) with hadronic initial state
at $T_i=200$ MeV.
}
\label{fighm}
\end{figure}

The above statement together with the uncertainty of the
 critical temperature $T_c$~\cite{latt} poses the following question: Is
 the  existence of the QGP phase essential to reproduce the 
 WA98 data? To study this problem, we have considered two 
  possibilities: (iii) pure hadronic model without medium-modifications,
   and (i) pure  hadronic model  with scaling hypothesis according to
Eq.(\ref{anst}) (for $\lambda=1/2$).
   In the former case,   
$T_i$ is found to be $\sim 250 $ MeV for 
$\tau_i=1$ fm/c and $dN/dy=700$,
 which appears to be too high for the hadrons to survive. Therefore
 we exclude this possibility here.
  On the other hand, 
 the second case with an assumption of $T_i = T_c$ (which is
  just for simplicity) leads to 
  $T_i\sim 205$ MeV, at $\tau_i= 1$ fm/c, which is not unrealistic.
In this case, the hadronic
system expands and cools and ultimately 
freezes out at $T_f$=120 MeV. 
The masses of the vector mesons increase 
with reduction in temperature (due to expansion) according to Eq.(\ref{anst}).
The results of this scenario  
for three values of the initial radial velocity including the prompt photon
contribution are shown in Fig.~(\ref{fighm}). 
The experimental data are well reproduced for vanishing 
initial transverse velocity also. Therefore, the results 
shown in Fig.~(\ref{figqgp}) and (\ref{fighm}) 
indicate that a simple hadronic model is inadequate.
 Either substantial modifications of hadrons in the thermal bath 
  or the formation of QGP in the initial stages is necessary to
   reproduce the data. It is rather difficult to distinguish 
between the two at present.    

\begin{figure}[htbp]
\epsfxsize=8cm
\centerline{\epsfbox{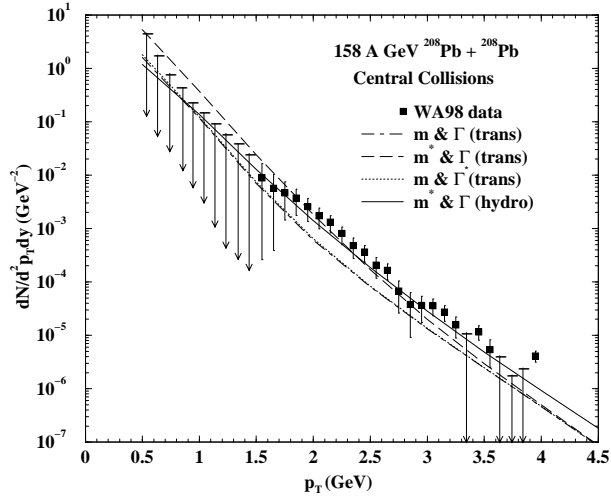}}
\caption{Same as Fig.~(\protect{\ref{fighm}}). Results for hydrodynamic
and transport (indicated by `trans') model calculations are compared. 
}
\label{figtrans}
\end{figure}

In Fig.~(\ref{figtrans}) we compare the results of hydrodynamic 
and transport calculations.  
Within the framework of the transport model 
the data is well reproduced 
when the hadronic masses 
are allowed to vary (scenario i) according to  the Eq.~\ref{anst} 
($\lambda=1/2$,long dashed line). 
As mentioned before the photon spectra is 
insensitive to the change in width (scenario ii).
In the scenario (ii) the experimentally observed ``excess'' photon 
in the region $1.5\le\,p_T$ (GeV) $\le 2.5$ (dotted line)
is not reproduced. 
The change in the width of the vector mesons ($\rho$ 
in particular) has very little effects on the $p_T$
spectra of photons due to the following reasons.  
The density of an unstable particle  in a
thermal bath can be written as~\cite{weldon},
\be
\frac{dN}{d^3kd^3xds}=\frac{g}{(2\pi)^3}e^{-\sqrt{k^2+s}/T}\,P(s)
\label{bolt}
\ee
where $g$ is the statistical degeneracy of the particle and $P(s)$ is
the spectral function,
\be
P(s)=\frac{1}{\pi}\frac{{\s {Im}}\,\Pi}{(s-m_\rho^2-{\s {Re}}\,\Pi)^2
+({\s {Im}}\,\Pi)^2}
\label{sfn}
\ee
${\s {Im}}\,\Pi$ (${\s {Re}}\,\Pi$) is the imaginary (real) part of the 
(trace of) $\rho$ self energy. 
Eqs.~(\ref{bolt}) and ~(\ref{sfn})
indicate that the density of particles in a thermal
bath is given by the Boltzmann distribution
weighted by the Breit-Wigner
function, which gets maximum weight from the 
value of $s=m_\rho^2+{\s {Re}}\,\Pi$,
the contribution
from either side of the maximum being averaged out. 
Therefore, the results become sensitive to the
effective mass, $s=m_\rho^{2*}=m_\rho^2+{\s {Re}\Pi}$ and not to
the width of the spectral distribution. 
The dash-dotted line indicates results
with vacuum masses and widths. 
In case of transport model calculations (dashed line)
there is excess photons at the low $p_T$
region compared to the hydrodynamic model (solid line) due
to the following reason. 
We find that the variation of temperature with time (cooling law)
in Eq.~\ref{trans} is slower than the one obtained by solving 
hydrodynamic equations. As a consequence the thermal system
has a longer life time than the former case, allowing the
system to emit photons for a longer time. 
In case of hydrodynamics this is compensated by the transverse kick 
experienced by the photon at large $p_T$ due to radial velocity of the 
expanding matter. 

It is widely believed that the photon spectra can be very useful
for the estimation of the initial temperature. In table I
We show the initial temperature obtained by several 
authors~\cite{ss,st,hrr,gk,pp} by analyzing the WA98 photon spectra.
In ~\cite{gk} the thermal photon spectra was parametrized as
$EdN/d^3p=V_4F(E,T_{eff},v_0)$, similar to the parametrization 
one uses to study the thermal hadronic spectra. A value of $T_{eff}=170$
MeV and $v_0=0.3$ can reproduce the WA98 data if the prompt photon contribution
is normalized to reproduce the data for $p_T>2$ GeV.
 As $T_{eff}$ is the average value of the temperature ($T_i>T_{eff}>T_f$)
in this analysis, one may expect an initial temperature $\sim 200$ MeV.
In ~\cite{pp} the hard photon contribution has been used to 
normalized the scaled p-p data at $\sqrt{s}=19$ GeV  and the thermal
photon spectra was evaluated for both QGP and hadronic initial state
assuming a non-zero radial velocity of the QGP/hadronic fluid. 
Steffen and Thoma~\cite{st} have demonstrated that the value 
of the statistical degeneracy ($g_h$) in the hadronic phase (and hence
the life time of the mixed phase) is crucial for the description of WA98 data.
An initial temperature $\sim 220$ MeV with $T_c=170$ MeV and
$g_h=8$ can reproduce the data in this case. In~\cite{ss} and
~\cite{st} the static photon emission rate from QGP is similar, then
why the value of $T_i$ is so different? Apparently the main reason
is the difference in the life time of the mixed phase between
the two cases. In~\cite{st} the life time of the mixed phase is
large (small hadronic degeneracy) compared  to that of ref.~\cite{ss},
hence allowing the mixed phase to emit photon for a longer time interval
in the former case. In~\cite{st} the contributions from the
QGP phase due to smaller initial temperature (compared to~\cite{ss})
is compensated by a larger contributions from the mixed phase.
Huovinen et. al.~\cite{hrr} has studied in detail the effects of various
EOS, boost invariant and non-invariant hydrodynamic flow on the
photon spectra. Thermal photon with values of $T_i\sim 213 - 255$ MeV
(depending on the EOS and evolution scenario)  
+ hard QCD photon can reproduce the data well in their case. We 
would like to mention here that the (thermal) photon emission 
rate used in~\cite{hrr} is full order of $\alpha_s$
results obtained in~\cite{pmy}. 

\begin{center}
\begin{tabular}{|c|c|c|c|c|c|}
\hline
~\cite{ss} & \cite{gk} & \cite{st} & \cite{pp} & \cite{hrr} & Present\\
\hline
$T_i$ & $T_i$ & $T_i$ &  $T_i$ & $T_i$ & $T_i$ \\
\hline
335 & 210 & 220 & 200-230 & 213-255 & 200 \\
\hline
\end{tabular}
\end{center}
\vskip 0.1in
\noindent{Table I: The initial temperature obtained
in the present analysis is compared with the values
obtained by other authors~\cite{ss,gk,st,pp,hrr}. 
}

In order to reproduce the WA98 photon data either a 
substantial reduction in vector meson masses or the formation
of QGP in the initial stage with $T_i\sim 200$ MeV
is necessary. A simple hadronic model is appears to
be ruled out by the experimental data.

In spite of the encouraging situation mentioned above, 
a firm conclusion about the formation of the
QGP at SPS necessitates a closer look at some pertinent
but unsettled issues. 
A prerequisite for the detection of the QGP by studying the photon spectra is to 
subtract contribution from the initial hard processes.
Therefore, it is extremely important
to know quantitatively the contribution from the hard processes.
Again, the assumption of complete thermodynamic equilibrium
for quarks and gluons may not be entirely
realistic for SPS energies; lack of chemical equilibrium of quarks
will further reduce the thermal yield from QGP. 
We have assumed $\tau_i=1$ fm/c at SPS energies, 
which may be considered as the
lower limit of this quantity,  because
the transit time (the time taken by the nuclei to pass
through each other in the CM system) is $\sim$ 1 fm/c at SPS
energies and
the thermal system is assumed to be formed after this time 
has elapsed. 
In the present work, when QGP initial state
is considered, we have assumed a first order phase transition
with bag model EOS for the QGP for its simplicity, although 
it is not in complete agreement with the lattice QCD 
simulations~\cite{latt}. However, it is difficult to distinguish
among different EOS with the current resolution of the photon data
~\cite{hrr}. 
As mentioned before, 
there are uncertainties in the value of $T_c$~\cite{latt},
a value of $T_c\sim 200$ MeV may be considered as an
upper limit.  Moreover, the photon
emission rate from QGP given by Eq.~(\ref{edr}), evaluated in 
Refs.~\cite{kapusta,aurenche}
by resumming the hard thermal loops is strictly valid for $g<<1$ 
whereas the value of $g$ obtained from Eq.~(\ref{strong}) is $\sim 2$
at $T\sim 200 $ MeV. At present it is not clear whether the rate
in Eq.~(\ref{edr}) is valid for such a large value of $g$ or not.
New method is required to evaluate the photon emission rate from QCD 
Plasma, HTL approximation is not valid at SPS energies and it may not
be valid even at LHC energies. Evaluation of the spectral function from
lattice QCD~\cite{nah} will be very useful in this context.

\section{Acknowledgement}
This work was done in collaboration with Sourav Sarkar, Tetsuo Hatsuda, 
Pradip Roy, Tapan K. Nayak and Bikash Sinha. I am grateful to all of them.

\end{document}